\documentclass[twocolumn,showpacs,preprintnumbers,prl,aps,amssymb,superscriptaddress]{revtex4}
\usepackage{graphicx}
\usepackage{dcolumn}
\usepackage{bm}
\begin{document}

\newcommand{\Co}{CeCoIn$_5$}
\newcommand{\Rh}{CeRhIn$_5$}
\newcommand{\ie}{{\it i.e.}}
\newcommand{\eg}{{\it e.g.}}
\newcommand{\etal}{{\it et al.}}

\title{ Nonvanishing Energy Scales at the Quantum Critical Point of \Co}

\author{Johnpierre~Paglione}
\affiliation{Department of Physics, University of Toronto, Toronto M5S 1A7, Canada}
\author{M.A.~Tanatar}
\altaffiliation[Permanent address: ] {Institute of Surface Chemistry, N.A.S. Ukraine, Kyiv 03164, Ukraine.}
\affiliation{Department of Physics, University of Toronto, Toronto M5S 1A7, Canada}
\author{D.G.~Hawthorn}
\affiliation{Department of Physics, University of Toronto, Toronto M5S 1A7, Canada}
\author{F.~Ronning}
\affiliation{Department of Physics, University of Toronto, Toronto M5S 1A7, Canada}
\author{R.W.~Hill}
\affiliation{Department of Physics, University of Toronto, Toronto M5S 1A7, Canada}
\author{M.~Sutherland}
\affiliation{Department of Physics, University of Toronto, Toronto M5S 1A7, Canada}
\author{Louis Taillefer}
\email{Louis.Taillefer@USherbrooke.ca}
\affiliation{Department of Physics, University of Toronto, Toronto M5S 1A7, Canada}
\affiliation{D\'epartement de physique and RQMP, Universit\'e de Sherbrooke, Sherbrooke J1K 2R1, Canada}
\author{C.~Petrovic}
\affiliation{Condensed Matter Physics Department, Brookhaven National Laboratory, Upton, New York 11973, USA}

\date{\today}

\begin{abstract}

Heat and charge transport were used to probe the magnetic field-tuned quantum critical point in the heavy-fermion metal CeCoIn$_5$. A comparison of electrical and thermal resistivities reveals three characteristic energy scales. A Fermi-liquid regime is observed below $T_{FL}$, with both transport coefficients diverging in parallel and $T_{FL}\to 0$ as $H\to H_c$, the critical field. The characteristic temperature of antiferromagnetic spin fluctuations, $T_{SF}$, is tuned to a minimum but {\it finite} value at $H_c$, which coincides with the end of the $T$-linear regime in the electrical resistivity. A third temperature scale, $T_{QP}$, signals the formation of quasiparticles, as fermions of charge $e$ obeying the Wiedemann-Franz law. Unlike $T_{FL}$, it remains finite at $H_c$, so that the integrity of quasiparticles is preserved, even though the standard signature of Fermi-liquid theory fails.

\end{abstract}

\pacs{72.15.-v,42.50.Lc,71.10.Hf,71.27.+a}

\maketitle

The ongoing search for universality in systems tuned to a quantum critical point (QCP) has led to the discovery of a host of fascinating condensed matter systems which deviate from Landau's Fermi liquid (FL) theory of metals. With dominant characteristic energy scales which become small or vanishing at a QCP, the Fermi energy no longer dictates the form of low-energy excitations, and so-called non-FL behaviour prevails \cite{Stewart}. 

The extent to which zero-temperature critical fluctuations influence the fermionic degrees of freedom at a QCP is an open question. For instance, two leading theories predict quite different fates for the FL state. In the weak-coupling quantum spin density wave (SDW) scenario \cite{HertzMillis,Moriya}, fluctuations are concentrated at hot spots on the Fermi surface, leading to a ``mild'' breakdown of FL theory: at the QCP, the electronic specific heat $C/T$ shows a square-root divergence but remains finite in the $T \to 0$ limit \cite{Moriya2}, reflecting the fact that, below a {\it finite} characteristic temperature, the FL state is recovered on part of the Fermi surface. This scenario appears to be realized in CeNi$_2$Ge$_2$ \cite{Gegenwart} and CeIn$_3$ \cite{CeIn3}, and is usually accompanied by a $T^{3/2}$  dependence of resistivity \cite{Moriya2}. In the strong-coupling ``locally'' critical scenario \cite{Si,Coleman}, fluctuations are thought to completely cover the Fermi surface, causing a logarithmic divergence of $C/T$ and a vanishing characteristic temperature \cite{Coleman}. This leads to a ``strong'' breakdown of the quasiparticle picture \cite{Coleman}. This scenario is thought to be realized in YbRh$_2$Si$_2$ \cite{Gegenwart,Custers} and CeCu$_{5.9}$Au$_{0.1}$ \cite{Schroder}, and is characterized by a $T$-linear resistivity at the QCP.

The comparison of heat and charge transport is one of a few experimental studies which can give access to information on the spectrum of critical fluctuations {\it and} their influence on fermionic excitations. A quintessential test of FL theory is the Wiedemann-Franz (WF) law, which states that the ratio of thermal ($\kappa$) to electrical ($\sigma$) conductivities is a universal constant in the $T \to 0$ limit: $\kappa / \sigma T = L_0 \equiv \frac{\pi^2}{3} \left(\frac{k_B}{e}\right)^{2}$. A violation of this law would imply a profound breakdown of the FL model, in the sense that low-lying excitations would no longer be quasiparticles of charge $e$ obeying Fermi statistics. In addition, a comparison of $\kappa(T)$ and $\sigma(T)$ at finite temperatures provides information about the momentum and energy dependence of magnetic fluctuations, through their effect on quasiparticle scattering, and thus can also be used to probe the nature of a QCP. 

In this Letter, we apply this approach to a system with tunable critical behaviour in order to (i) test the WF law at the QCP and (ii) track the fluctuation spectrum as a function of tuning parameter. The material, \Co, is a heavy-fermion metal which exhibits a magnetic field-tuned QCP characterized by a divergence in transport \cite{Paglione} and thermodynamic \cite{Bianchi} quantities at a critical field $H_c$. With a readily accessible and continuous control parameter, this extremely clean, stoichiometric material offers a unique opportunity to study criticality via heat transport over the entire temperature range of relevance.

Heat and charge transport measurements were performed as described previously \cite{Paglione_Rh,Tanatar} on single crystals of \Co\ grown by the self-flux method \cite{Cedomir} with $\rho_0 \simeq 0.1~\mu\Omega$~cm ($H \to 0$), for currents parallel to [100] and field parallel to [001]. A comparison of heat and charge resistivities reveals that scattering in \Co\ is practically identical to that observed in antiferromagnetic \Rh\ above its ordering temperature, $T_N$ \cite{Paglione_Rh}, both being governed by a comparable spin-fluctuation scale $T_{SF}$. This confirms the magnetic nature of the QCP in \Co\ \cite{magnetic_pressure}. Moreover, in \Co\ $T_{SF}$ is tuned by magnetic field towards a minimum but {\it finite} value at $H_c$, which accounts for the departure at low $T$ from the $T$-linear resistivity. We show that, despite the presence of a non-FL $T^{3/2}$ power law in both electrical and thermal resistivities at $H_c$, the WF law is still obeyed in the $T \to 0$ limit. This reveals a ``mild'' breakdown of FL theory in \Co\, consistent with the SDW-model.

\begin{figure}
\centering
\includegraphics[totalheight=2.5in]{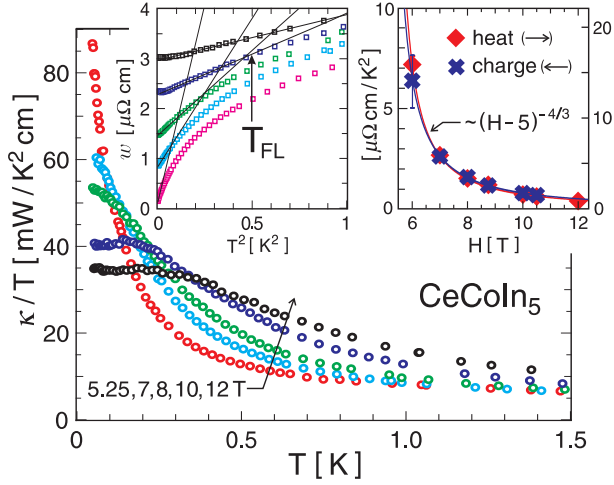}
\caption{\label{fig:Kappa} 
Thermal conductivity of \Co, plotted as $\kappa/T$ vs $T$ (main panel) and as electronic \cite{phonon} thermal resistivity $w=L_0T/\kappa_e$ vs $T^2$ (left inset), for $H\parallel [001]$. The data in the left inset, offset for clarity, is for $H = 6$, 7, 8, 10 and 12~T (bottom to top); lines are linear fits valid up to $T = T_{FL}$, the Fermi-liquid temperature, marked by an arrow for $H=10$~T. 
Right inset: field dependence of the $T^2$ Fermi-liquid coefficients of charge and heat transport.}
\end{figure}

\begin{figure}
\centering
\includegraphics[totalheight=3.5in]{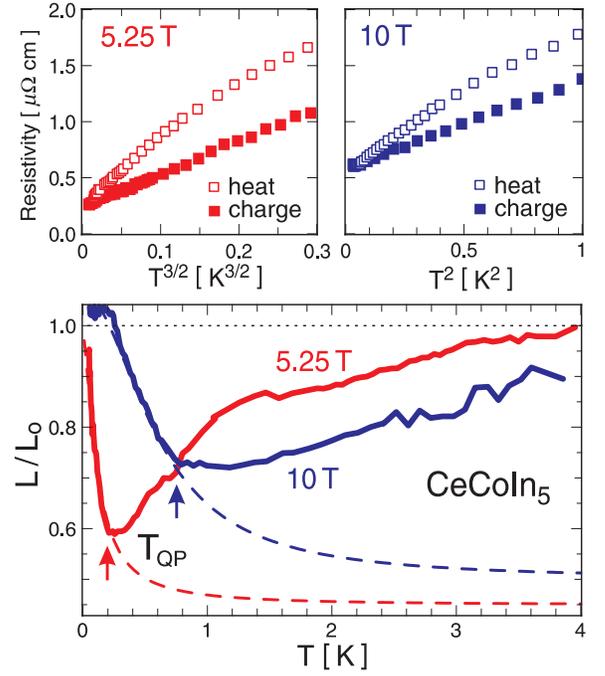}
\caption{\label{fig:w_vs_rho} 
Upper panels: comparison of thermal ($w(T)$; open symbols) and electrical ($\rho(T)$; solid symbols) resistivities at the critical field (5.25~T), plotted vs $T^{3/2}$ (left), and in the FL regime (at 10~T),  plotted vs $T^2$ (right). 
Lower panel: normalized Lorenz ratio, $L/L_0 \equiv \kappa_e/ L_0 \sigma T \equiv \rho(T)/w(T)$, vs $T$. Dashed lines show the ratio of the low-temperature power laws, namely $(\rho_0 + a T^{\alpha}) / (w_0 + bT^{\alpha})$, with $\alpha = 3/2$~and 2, for 5.25 and 10 T, respectively. 
The quasiparticle temperature, $T_{QP}$, marked by arrows, is defined as the temperature below which $L(T)$ starts to rise, aiming towards unity.}
\end{figure}

\begin{figure}
\centering
\includegraphics[totalheight=2.5in]{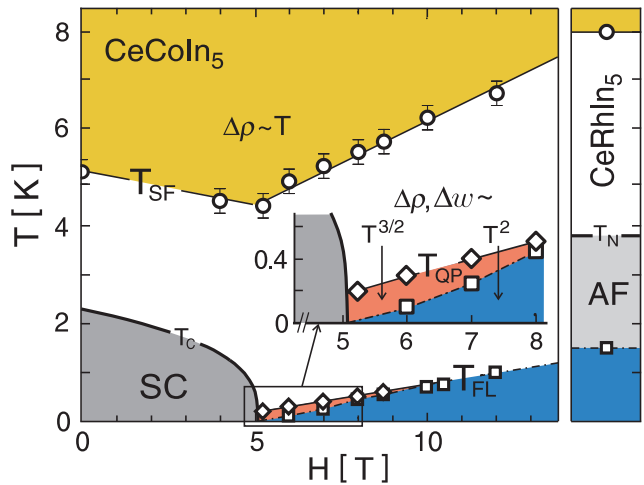}
\caption{\label{fig:PD} Evolution of characteristic energy scales in \Co\ vs magnetic field. The Fermi-liquid temperature $T_{FL}$ is the end of the $T^2$ regime in $w(T)$ (squares). The quasiparticle temperature $T_{QP}$ is the onset of the low-$T$ upturn in $L(T)$ (diamonds). The spin-fluctuation temperature $T_{SF}$ is reached when $\delta (T)=0$ at high $T$ (circles). Error bars for $T_{QP}$ and $T_{FL}$ are smaller than the size of symbols. Note that $T_{QP} = T_{FL}$ at $H=10$~T and above. To the right, we also show $T_{SF}$ and $T_{FL}$ for \Rh\ (at $H=0$).}
\end{figure}

{\it Fermi-liquid temperature, $T_{FL}$.} Previous resistivity measurements \cite{Paglione} have shown that a FL regime develops in \Co\ above its superconducting $H_{c2} = 5$~T, characterized by $\Delta\rho \equiv \rho-\rho_0 = AT^2$, with $A$ diverging as $A(H)\propto (H-H_{c,A})^{-\alpha}$, where $H_{c,A}= 5.1$~T and $\alpha = 1.37 \approx 4/3$. Fig.~\ref{fig:Kappa} presents an analysis of $\rho(T)$ and $\kappa(T)$ data obtained from a new sample with resistivity characterized by very similar fit parameters, namely $H_{c,A}=5.0\pm0.1$~T and $\alpha=1.29\pm0.1$. As a function of field, $\kappa/T$ evolves from an almost divergent behaviour at 5.25~T towards more FL-like saturation at higher fields. This is seen more clearly by plotting the electronic \cite{phonon} thermal resistivity $w\equiv L_0 T/ \kappa_e$ vs $T^2$, in the left inset of Fig.~\ref{fig:Kappa}. This plot reveals a $T^2$ dependence of $w(T)$ (\ie\ $\Delta w \equiv w-w_0 = BT^2$), observed below a characteristic temperature $T_{FL}$ as high as 1.0 K at 12 T, which decreases steadily, so that $T_{FL} \to 0$ at $H_c$ (see Fig.~\ref{fig:PD}). 

The field dependence of the slope $B$, which represents the contribution of electron-electron (e-e) scattering to thermal transport (analogous to $A$), is shown in the right inset of Fig.~\ref{fig:Kappa}, together with $A(H)$. It is clear that $B(H)$ has the {\it same critical field dependence} as $A(H)$. Specifically, $B$ is best fitted by a function $B(H)\propto (H-H_{c,B})^{-\beta}$ with parameters $ H_{c,B}=5.0\pm0.2$ ~T and $\beta=1.34\pm0.1$, so that $ H_{c,A}= H_{c,B} \equiv H_c = 5.0$~T and $\alpha = \beta$ (within error). Therefore, $A(H)$ and $B(H)$ differ only by a {\it field-independent} factor, $A/B = 0.47$ $\pm0.03$. Since the ratio $A/B$ is governed by the {\bf q}-dependence (\ie\ is sensitive to the angular dependence) of e-e scattering around the Fermi surface \cite{Bennett}, this suggests that the anisotropy of quasiparticle scattering in \Co\ is unchanged by the field, even though $A$ itself grows by a factor of 35, from $0.2~\mu\Omega$~cm/K$^2$ at 16~T to $7~\mu\Omega$~cm/K$^2$ at 6~T.  

{\it Quasiparticle temperature, $T_{QP}$.} As we approach the QCP, the ranges of $T^2$ thermal and electrical resistivities shrink to nothing (\ie\ $T_{FL} \to 0$), whereupon both $\Delta \rho$ and $\Delta w$ exhibit a different power-law dependence at low temperature, namely $T^{3/2}$ (see upper panels of Fig.~\ref{fig:w_vs_rho}).  Remarkably, the $T \to 0$ extrapolations of $\rho(T)$ and $w(T)$ {\it within this non-FL regime} nevertheless converge to satisfy the WF law, so that $\rho_0 = w_0$ (within the $\pm$6\% experimental accuracy on the ratio) not only far from $H_c$  (\eg\ at 10~T) but also right at $H_c$ (\ie\ at 5.25~T). This reveals that the breakdown of FL theory in \Co\ is not complete: while the expected $T^2$ dependence of the scattering rate is indeed violated at the QCP, {\it the integrity of the quasiparticles themselves is nonetheless preserved}.

The overall temperature dependence is best captured by plotting the (normalized) Lorenz ratio, $L/L_0 \equiv \kappa_e/ L_0 \sigma T \equiv \rho(T)/w(T)$, shown in the lower panel of Fig.~\ref{fig:w_vs_rho}. The convergence of $\rho(T)$ and $w(T)$ shows up as a rapid upturn in $L/L_0$ with decreasing $T$, wherefrom it is aimed at unity. We define as $T_{QP}$ the onset of this upturn, which is also the temperature below which $\Delta \rho$ and $\Delta w$ have both reached their asymptotic power-law behaviour. {\it We view $T_{QP}$ as the temperature below which quasiparticles form}. In the inset of Fig.~\ref{fig:PD}, we plot $T_{QP}$ as a function of field. Away from the QCP, for $H \geq 10$~T, $T_{QP}$ coincides with $T_{FL}$, so that  quasiparticles exhibit the standard $T^2$  behaviour as they form. However, as one approaches the QCP, for $H < 10$~T, the upturn in $L(T)$ starts above $T_{FL}$, and $T_{QP}$ remains finite as $T_{FL}$ vanishes. Therefore, quasiparticles still form at the QCP of \Co, even though they do not show the standard FL signature of $T^2$ resistivity. This is reminiscent of the observation that quantum oscillations are still present at $H_c$, while standard Lifshitz-Kosevich theory fails \cite{Alix}. For a complete breakdown of quasiparticles at the QCP, one would need to have seen $T_{QP} \to 0$, in addition to the usual condition $T_{FL} \to 0$. It transpires that {\it $T_{QP}$ is a new and fundamental temperature scale for quantum criticality}.

\begin{figure}
\centering
\includegraphics[totalheight=2.5in]{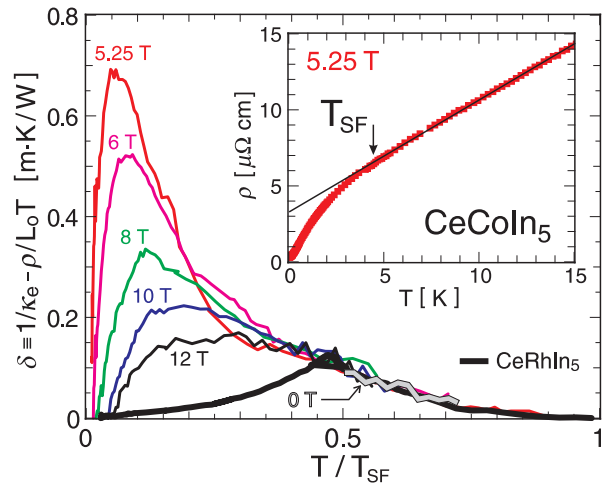}
\caption{\label{fig:delta} Difference between electronic thermal resistivity and electrical resistivity of \Co, labelled $\delta$, as a function of reduced temperature $T/T_{SF}$, where $T_{SF}$ is obtained by making all curves at different fields (as indicated) match at high temperature.  The corresponding data for antiferromagnetic \Rh\ in zero field is also shown (thick black line). Inset: temperature dependence of electrical resistivity at the critical field, with arrow indicating the position of $T_{SF}$. The line is a linear fit to the data above 8~K.}
\end{figure}

{\it Spin fluctuation temperature, $T_{SF}$}. A recent study of \Rh\ \cite{Paglione_Rh} has shown the usefulness of examining not only the ratio, but also the {\it difference} between thermal and electrical resistivities, given by $\delta \equiv 1/\kappa_e - \rho/L_0T$ \cite{delta}. Physically, $\delta(T)$ is due to those scattering processes in which the energy of the conduction electron is changed but not its direction, thereby affecting $\kappa$ but not $\sigma$ \cite{Kaiser}. It is plotted in Fig.~\ref{fig:delta} for both \Co\ and \Rh. We begin by describing its behaviour in \Rh, where the electronic scattering rate was observed to be directly proportional to the entropy of the magnetic system (specifically, $w \propto S_{mag}$, the magnetic entropy)~\cite{Paglione_Rh}. 

At high temperature, $\delta(T)$ goes to zero, not because the overall scattering has decreased, for that keeps increasing monotonically with $T$ (tracking $S_{mag}$), but because direction-conserving processes have become ineffective. This occurs when $T$ exceeds the characteristic temperature $T_{SF}$ of spin fluctuations, which then have insufficient energy to scatter electrons through the thermal layer \cite{Kaiser}. We define $T_{SF}$ to be the temperature where $\delta(T) \to 0$ \cite{Debye}. In \Rh, $T_{SF} \simeq 8$~K \cite{Paglione_Rh}, the temperature where, interestingly, neutron studies found antiferromagnetic correlations to set in \cite{Bao}. As temperature is decreased below $T_{SF}$, $\delta(T)$ starts to rise, and keeps rising until $T_N$, where it takes an abrupt cusp-like dive, as a gap opens in the fluctuation spectrum upon ordering. At $T$ well below $T_N$ the electron system eventually enters a FL state characterised by: (1) a linear rise  in  $\delta(T)$ up to $T_{FL} \simeq 1.5$~K, from the $T^2$ dependence of both $w$ and $\rho$; (2) a low mass enhancement, with $A = 0.02 ~\mu \Omega$~cm/K$^2$ \cite{Paglione_Rh}, and (3) the WF law, $\delta(T) \to 0$ at $T \to 0$.

Turning to \Co, one can see from Fig.~\ref{fig:delta} that at high temperature $\delta(T)$ curves for all fields can be collapsed onto the $\delta(T)$ curve for \Rh\ above $T_N$, upon normalizing $T$ by $T_{SF}$. The values of $T_{SF}$ needed for this scaling are plotted in Fig.~\ref{fig:PD}. By inspection, one can see that \Co\ at $\simeq 15$~T is equivalent to \Rh\ for $T > 4$~K, in the sense that the two materials have the same $\rho(T)$ and $\delta(T)$, the same $T_{SF} \simeq 8$~K and even the same $T_{FL} \simeq 1.5$~K. Therefore, {\it the electrons are scattered by the same antiferromagnetic fluctuations in both materials}. The difference occurs below 4~K: while the magnetic moments in \Rh\ order at $T_N = 3.8$~K, they never do in \Co\ where the entropy remains high all the way down to the FL regime, leading to a large mass enhancement, with a coefficient $A = 0.2~\mu\Omega$~cm/K$^2$ (at 16~T) \cite{Paglione}, 1 order of magnitude larger than in \Rh. 

As the field is decreased towards $H_c$, $T_{SF}$ steadily drops towards a minimum value of 4.4~K at $H_c$ (see Fig.~\ref{fig:PD}). This shows that the antiferromagnetic fluctuations in \Co\ are indeed tuned by the magnetic field. Note, however, that $T_{SF}$ does not vanish at $H_c$. This fact is an important new element in our understanding of quantum criticality in \Co. In particular, it elucidates why the resistivity does not display a single power law at the QCP: as shown in the inset of Fig.~\ref{fig:delta}, $\rho(T)$ at $H_c$ is linear down to 5~K, but then drops as it crosses $T_{SF}$, to eventually go over to a $T^{3/2}$ dependence.

A finite $T_{SF}$ suggests that the energy of magnetic fluctuations remains finite even at the QCP, as found with neutrons in CeNi$_2$Ge$_2$ \cite{Kadowaki}, which also obeys the WF law \cite{Kambe} as does Sr$_3$Ru$_2$O$_7$ at its field-tuned QCP \cite{Sr327}. Together with a $T^{3/2}$ dependence at $H_c$, also observed in CeIn$_3$ under both applied field \cite{CeIn3} and pressure \cite{Mathur}, and in \Co\ under pressures that restore the Fermi-liquid state at low $T$ \cite{Sidorov}, this is consistent with gaussian-type fluctuations predicted in the SDW model  \cite{HertzMillis,Moriya,Moriya2}. This presumably indicates that magnetic fluctuations in the CeIn$_3$ planes of \Co\ and in bulk CeIn$_3$ itself have a similar character.

Summarizing our observations, we can state that
(1) fluctuations near the field-tuned QCP in \Co\ are antiferromagnetic in nature, as revealed by the scaling of $\delta (T)$ curves for \Co\ relative to \Rh;
(2) the characteristic temperature scale of fluctuations, $T_{SF}$, is tuned to a minimum but non-vanishing value at the QCP;
(3) $T_{SF}$ correlates well with the end of the $T$-linear regime in the electrical resistivity, thus accounting for the lack of a single power law in $\rho(T)$ at the QCP;
(4) at $H_c$, both electrical and thermal resistivities exhibit a $T^{3/2}$ dependence below a second non-vanishing characteristic temperature, $T_{QP}$;
(5) even in the presence of such non-FL behaviour, the Wiedemann-Franz law holds in the $T\to 0$ limit at $H_c$.

These findings point to a ``mild'', incomplete breakdown of Fermi-liquid theory in \Co, characterized by a non-vanishing $T_{QP}$, the temperature below which fermionic quasiparticles of charge $e$ appear to still form, even at the QCP where the usual $T^2$ Fermi-liquid regime has shrunk to nothing ($T_{FL} \to 0$). This seems to be in line with the spin-density wave scenario of quantum criticality, even though it requires that the Kondo temperature, effectively removing local moments from the problem, be higher than $T_{SF}\simeq4$~K, and the single-impurity Kondo temperature in dilute Ce$_{1-x}$La$_x$CoIn$_5$ alloys was determined to be $\sim 1.5$~K \cite{Satoru1}.

This work was supported by the Canadian Institute for Advanced Research and a Canada Research Chair (L.T.) and it was funded by NSERC. It was partially carried out at the Brookhaven National Laboratory, which is operated for the U.S. Department of Energy by Brookhaven Science Associates (DE-Ac02-98CH10886). The authors gratefully acknowledge K.~Behnia, P.~C.~Canfield, N.~Harrison, Y.~B.~Kim, C.~P\'epin, A.~Rosch and M.~F.~Smith.



\begin{references}
\bibitem{phonon} Apart from the main panel of Fig.1, where raw data is shown, we always use the {\it electronic} component $\kappa_e = \kappa - \kappa_p$, obtained by subtracting a phonon contribution as described previously \cite{Paglione_Rh}. Note that $\kappa_p$ is negligible below $\sim 1$~K.
\bibitem{delta} This definition of $\delta$ differs from that of \cite{Paglione_Rh} by an overall factor of $T$, but it now conforms to that used in theoretical calculations, such as in \cite{Kaiser}.
\bibitem{Debye} $T_{SF}$ is a characteristic temperature analogous to the Debye temperature $\Theta_D$ in metals with dominant electron-phonon scattering, where indeed $\delta(T) \to 0$ at $T \simeq \Theta_D$.
\end{references}
\end{document}